\documentclass[runningheads]{llncs}
\usepackage{graphicx}
\usepackage[T1]{fontenc}

\begin{document}
\title{Less is More - diveXplore 5.0 at VBS 2021}
\author{Andreas Leibetseder\orcidID{0000-0002-9535-966X} \and 
 Klaus Schoeffmann} %besser weg: \orcidID{0000-0002-9218-1704}

\authorrunning{A. Leibetseder et al.}
\institute{Klagenfurt University, Institute of Information Technology (ITEC), \\ Klagenfurt, Austria
\email{\{aleibets,ks\}@itec.aau.at}}
\maketitle              % typeset the header of the contribution
\begin{abstract}

As a longstanding participating system in the annual Video Browser Showdown (VBS2017-VBS2020) as well as in two iterations of the more recently established Lifelog Search Challenge (LSC2018-LSC2019), diveXplore is developed as a feature-rich \textit{Deep Interactive Video Exploration} system. After its initial successful employment as a competitive tool at the challenges, its performance, however, declined as new features were introduced increasing its overall complexity. We mainly attribute this to the fact that many additions to the system needed to revolve around the system's core element -- an interactive self-organizing browseable featuremap, which, as an integral component did not accommodate the addition of new features well. Therefore, counteracting said performance decline, the VBS 2021 version constitutes a completely rebuilt version 5.0, implemented from scratch with the aim of greatly reducing the system's complexity as well as keeping proven useful features in a modular manner.

% From VBS2020
%Having participated in the three most recent iterations of the annual Video Browser Showdown (VBS2017-VBS2019) as well as in both newly established Lifelog Search Challenges (LSC2018-LSC2019), the actively developed \textit{Deep Interactive Video Exploration (diveXplore)} system combines a variety of content-based video analysis and processing strategies for interactively exploring large video archives. The system provides a user with browseable self-organizing feature maps, color filtering, semantic concept search utilizing deep neural networks as well as hand-drawn sketch search. The most recent version improves upon its predecessors by unifying deep concepts for facilitating and speeding up search, while significantly refactoring the user interface for increasing the overall system performance.

\end{abstract}


\section{Introduction}

% VBS 2020
% Developed by the Multimedia Information Systems group of Klagenfurt University's Institute of Information Technology (ITEC), diveXplore (Deep Interactive Video Exploration) is an iteratively evolving interactive video retrieval system originally designed for the annual Video Browser Showdown~\cite{VBS2014,Lokoc2018,lokoc2019interactive} (VBS). It is intended for enabling quickly searching large video datasets such as the competion's current V3C1 collection~\cite{berns2019v3c1} (approx. 1\,000\,h of video) -- the first part of a three-part Vimeo clip compilation labeled V3C~\cite{rosetto2018} (approx. 3\,800\,h of video). While operating on videos, diveXplore was employed in past VBS iterations (VBS2017~\cite{schoeffmann2017collaborative}, VBS2018~\cite{primus2018itec,leibetseder2018sketch}, VBS2019~\cite{schoeffmann2019autopiloting}) but as well specifically modified to \textit{lifeXplore} for taking part in the most recent two image-based Lifelog Search Challenge runs (LSC2018~\cite{Munzer2018}, LSC2019~\cite{leibetseder2019lifexplore}) -- an interactive competition aimed at determining the best systems and approaches for fast moment retrieval in lifelogging data~\cite{gurrin2019invited}. Being able to participate in such different competitions providing distinct kinds of data with a mere moderate amount of changes to the system, shows its versatility for multimedia retrieval in general, which as well becomes apparent when regarding the multitude of browsing and search features offered by diveXplore.

The annual video retrieval challenge known as the Video Browser Showdown~\cite{VBS2014,Lokoc2018,lokoc2019interactive,rossetto2020interactive} (VBS) as well as the newly developed Lifelog Search Challenge~\cite{gurrin2020introduction,gurrin2019invited} (LSC) based on personal lifelog data are team competitions that involve quick interactive search of large multimodal databases. As such they include multiple tasks, such as Known-Item Search (KIS) along with Ad-hoc Video Search\footnote{\url{https://www-nlpir.nist.gov/projects/tv2018/Tasks/ad-hoc/}} (AVS) as well as two competition modes featuring experts and novices using the various competing systems. These conditions, coupled with a relatively tight per-task time-limit of merely a few minutes, in general imposes two requirements on a participating system: the need to be as sophisticated as possible in order to retrieve meaningful results, while at the same time being as simple to use as possible for accommodating novice users familiar with neither the tasks nor the systems.

\begin{figure}[htbp!]
	\center
	\includegraphics[width=\textwidth]{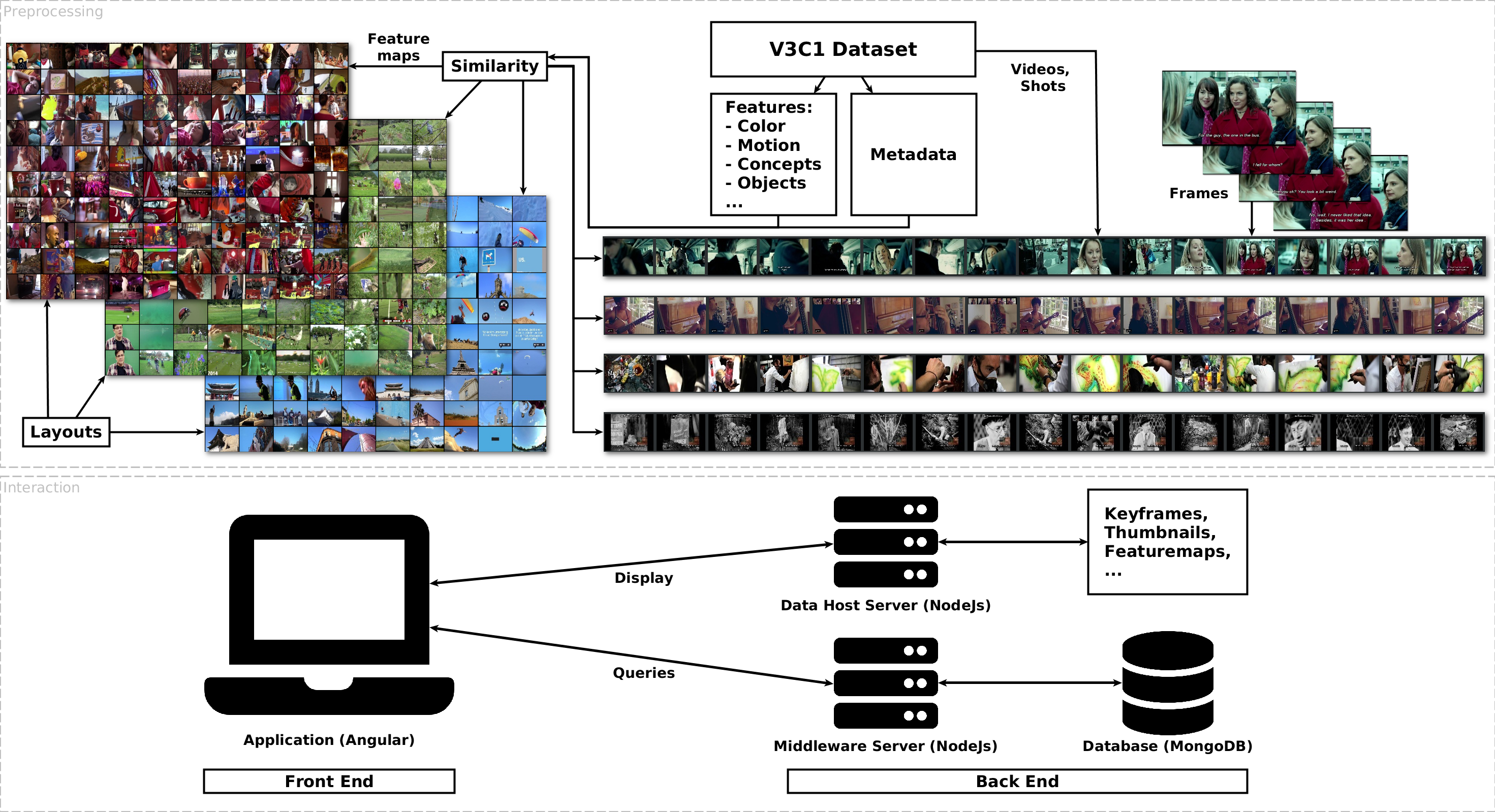}
    \caption{Architecture and of \textit{diveXplore} 5.0}
    \label{fig:architecture}
\end{figure}

The diveXplore system~\cite{schoeffmann2017collaborative,primus2018itec,leibetseder2018sketch,schoeffmann2019autopiloting,diveXplore2020} continuously participated in the VBS challenge since its 5th iteration (VBS2017-VBS2020) and grew in complexity along with the size of the competition's gradually growing dataset~\cite{rosetto2018}, which currently comprises approximately 1000 hours of video~\cite{berns2019v3c1}. Initially developed as a video exploration tool utilizing the concept of self-organizing feature maps~\cite{barthel2016navigating}, many new and exploratory features were added to the system for each VBS iteration, which led to an increasing deviation from its original core feature: video exploration through browsing. Consequently, the system became hard to maintain and complex to alter. Therefore, in order to competitively participate in VBS2021, diveXplore 5.0 is rebuilt from scratch focusing on the integration of successful features from the past challenges, while still offering video exploration features at a more modular level, preventing unnecessary feature dependencies. Figure~\ref{fig:architecture} portrays the system's new architecture, technology and features. The remaining sections describe diveXplore in detail as well as highlight novel additions to the system.

\section{diveXplore 5.0}

As a veteran system at the VBS, diveXplore was first developed for VBS2017~\cite{schoeffmann2017collaborative} incorporating an approach similar to the winning system at VBS2016~\cite{barthel2016navigating} for its main interface: an interactive, collaborative and self-organizing featuremap. Users were able to organize this map according to several similarity features, such as deep concepts, color, texture and motion. Additionally, these features could be utilized for similarity search on a keyframe basis with results being shown in separate views. Over the course of the last VBS iterations, these views became more and more important, as new features were introduced, such as a shot-aware video player, storyboards, color filter and textual concept as well as hand-drawn sketch search. All of these components needed to be integrated in accordance with the ever-present featuremap, such that navigating shots or videos in any of the views required re-positioning the currently displayed featuremap or even loading a new one. This overall slowed system performance in addition to being very error-prone. Therefore, the focus of diveXplore 5.0 is to remove these strong dependencies and construct separate views that do not have to be updated during most interactions.

\subsection{Architecture}

As illustrated in Figure~\ref{fig:architecture}, the system builds on several offline preprocessing steps in order to facilitate smooth interactive retrieval during the competition. First, the V3C1 dataset is split into custom shots as well as uniformly sampled frames using a one-second interval. This makes video retrieval very flexible since query results can be ordered on different levels of granularity. Subsequently, all included analyses can be started, which mostly include feature extraction utilizing hand-crafted as well as deep neural network approaches. The technologies used for creating the systems are web-based. The back end comprises two NodeJs\footnote{\url{https://nodejs.org}} servers, where one is used for hosting keyframes, thumbnails and featuremaps, while the other server acts as a RESTful middleware between a client written in Angular~\footnote{\url{https://angular.io}} and a MongoDB~\footnote{\url{https://www.mongodb.com}} database.

\subsection{Features}

As mentioned, diveXplore 5.0 keeps the most useful features from previous VBS iterations, while adjusting them to accommodate aforementioned analyses on shot as well as frame level. Consequently, it still provides a textual concept search that allows for the search of different deep concepts such as objects, attributes, locations as well as metadata. Furthermore, the possibility for similarity search is maintained as well, especially since it proved invaluable for AVS tasks: a user can select a particular shot or frame and can retrieve similar items by choosing a similarity measure such as deep concepts, color, texture or motion features. In addition, the system includes two additional explorative views that borrow from previously integrated components, yet incorporate novel strategies. The following sections describe these new features in more detail -- concept-based featuremaps and a video-based similarity filter.

\begin{figure}[htbp!]
	\center
	\includegraphics[width=\textwidth]{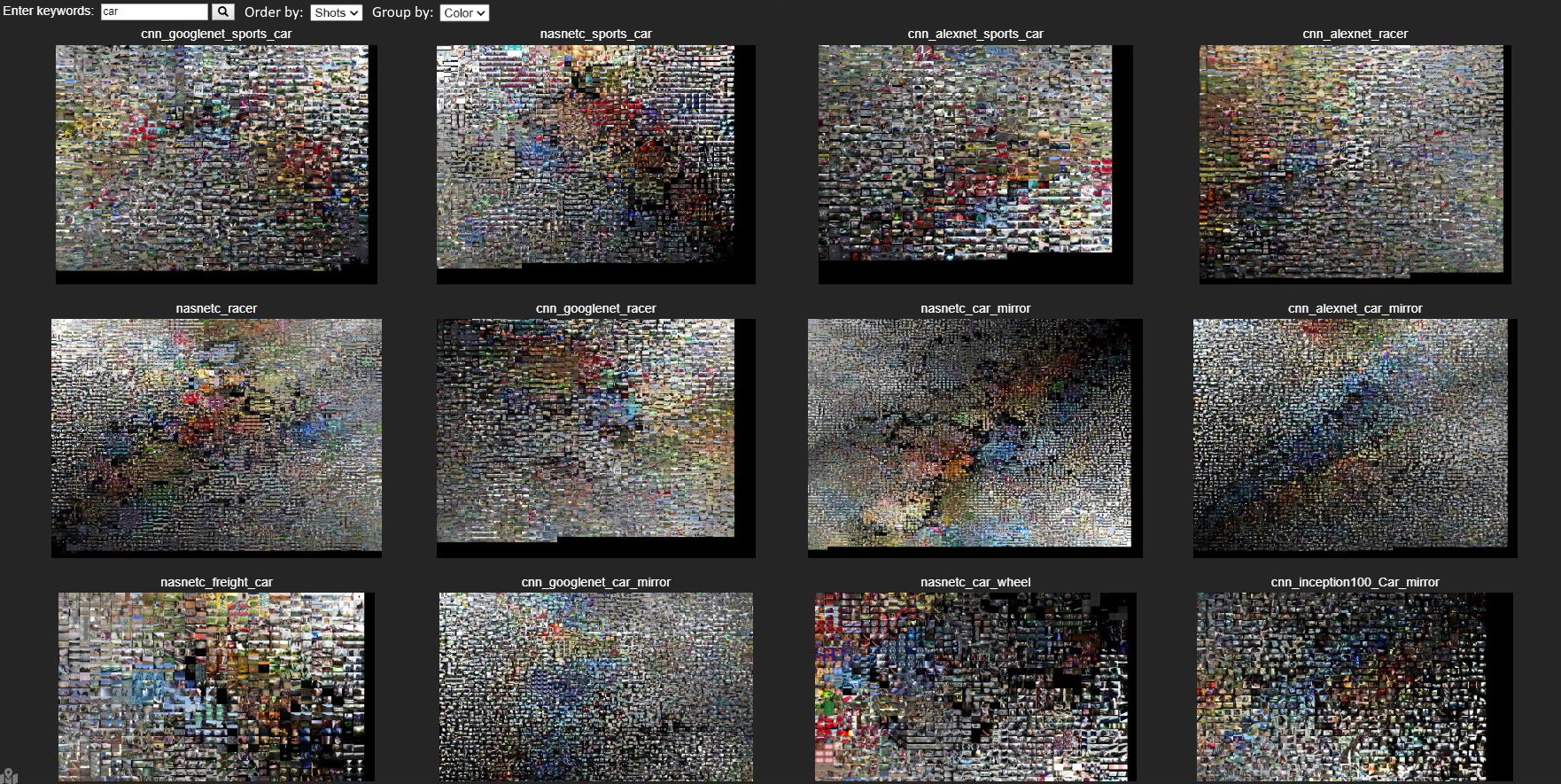}
    \caption{Concept-based featuremaps.}
    \label{fig:concept_featuremaps}
\end{figure}

\subsubsection{Concept-based Featuremaps}

Figure~\ref{fig:architecture} already indicates that diveXplore 5.0 keeps the concept of featuremaps, albeit in a different form: the large map incorporating the shots of all videos organized by different concepts is removed while maintaining and organizing smaller maps for individual concepts such as objects, which is illustrated in Figure~\ref{fig:concept_featuremaps}. By entering valid concepts, such as 'car' in the example, users can retrieve one or more featuremaps depending on how many incorporated deep nets include this particular concept. These smaller maps can subsequently be organized by utilizing all measures available for similarity search as well as recognition confidence and video affiliation. This feature is available for the multi-map overview shown in Figure~\ref{fig:concept_featuremaps} as well as an individual fullscreen view after selecting an individual featuremap.

\begin{figure}[htbp!]
	\center
	\includegraphics[width=\textwidth]{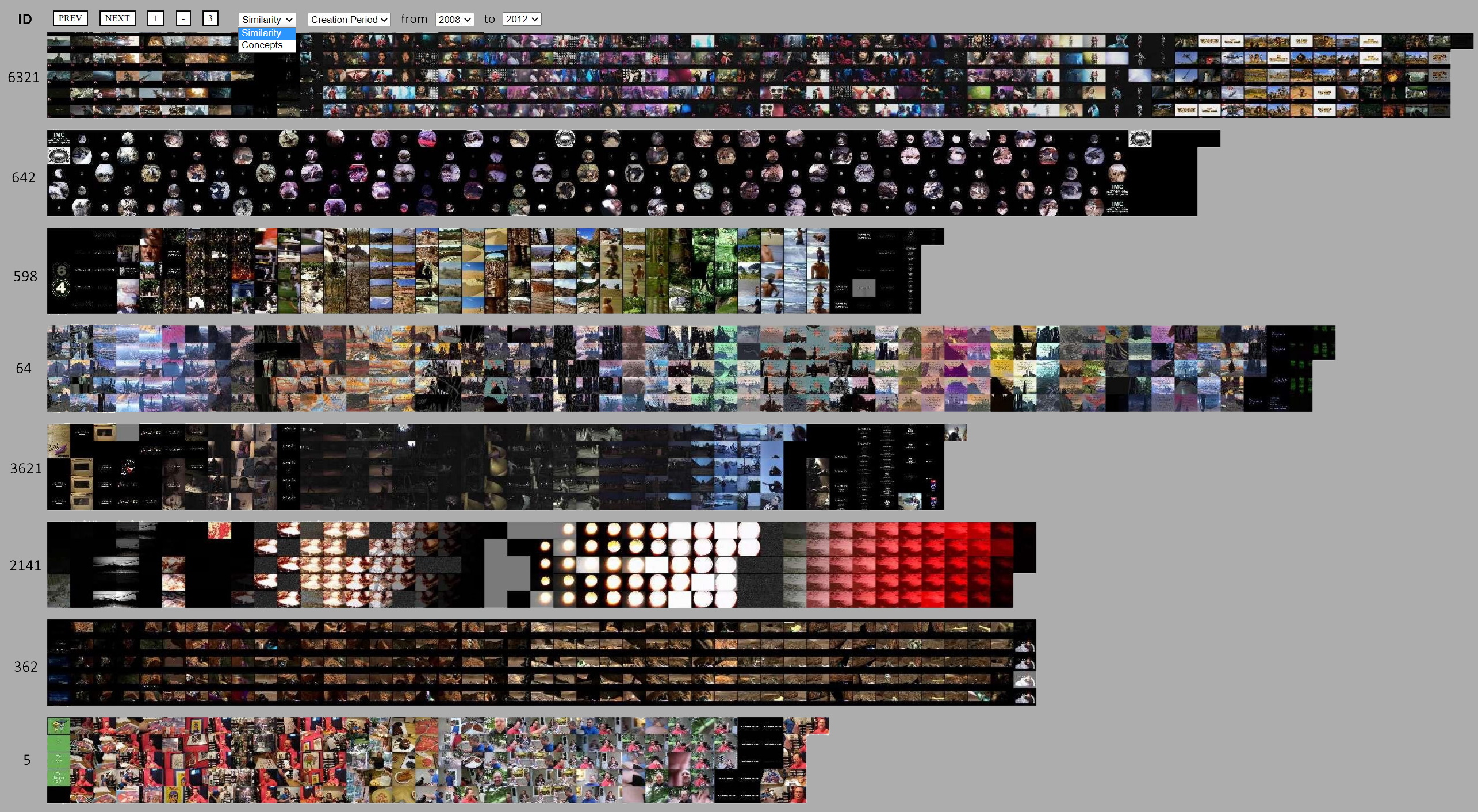}
    \caption{Video-based similarity filter.}
    \label{fig:video_similarity}
\end{figure}

\subsubsection{Video-based Similarity Filter}

The video-based similarity filter, shown in Figure~\ref{fig:video_similarity}, is a new addition based on the system's previously integrated storyboard feature, which provided a user with small summaries of each individual video. The main difference to the former implementation is the addition of interaction through several combinable filters. The view as well makes use of deep concepts and similarity, but for comparing entire videos or sections of them. As an example, the figure shows a situation where full videos are ordered by their creation period, i.e. between 2008 and 2012. Similarly, a user can as well select one or multiple concepts such as 'person' and 'apple' and group video sections according to either the frequency or classification confidence of these terms. Since this view allows for a multitude of different combinations it will expectedly be subject to many improvements in future system updates.

\section{Conclusion}

We present diveXplore 5.0 -- a completely rebuilt version of a long-time competing system at the annual Video Browser Showdown. While keeping some proven useful features, such as concept and similarity search, the system includes two additional views for dataset exploration: concept-based featuremaps enabling the exploration and organization of individual concepts, as well as a video-based similarity filter, which allows for video- or segment-wide similarity search according to various combinable criteria. All features are integrated on a shot- as well as uniformly sampled frame-basis in order to increase the system's flexibility. We expect these compared to diveXplore 4.0 much reduced and more carefully integrated components to shape a competitive system for VBS2021.

\section*{Acknowledgments}
This work was funded by the FWF Austrian Science Fund under grant P 32010-N38.

% \begin{figure}
% \includegraphics[width=\textwidth]{fig1.eps}
% \caption{A figure caption is always placed below the illustration.
% Please note that short captions are centered, while long ones are
% justified by the macro package automatically.} \label{fig1}
% \end{figure}

\bibliographystyle{splncs04}
\bibliography{refs}

\end{document}